\documentclass[letterpaper, 10 pt, conference]{ieeeconf}  % Comment this line out if you need a4paper

\IEEEoverridecommandlockouts                              % This command is only needed if 
\usepackage{amsmath} % assumes amsmath package installed
\usepackage{amssymb}  % assumes amsmath package installed
\usepackage{caption}
\usepackage{subcaption}
\usepackage{graphicx}
\usepackage{algorithm}
\usepackage{algpseudocode}
\usepackage[utf8]{inputenc}
\usepackage{hyperref}
\usepackage{subcaption}
\usepackage{multirow}
\usepackage{multicol}
\usepackage{cite} % For citation commands
\usepackage{tikz}
\usepackage{tikz-3dplot}
\usepackage{subcaption}
\usepackage{tabularx}
\usepackage{booktabs}    % for cleaner table lines
\usepackage{array}       % for more column control
\usepackage{epstopdf}

\providecommand{\keyword}[1]
{
  \small	
  \textbf{\textit{Keywords---}} #1
}

\title{\LARGE \bf
Powered Descent Trajectory Optimization of Chandrayaan-3 using Radau Collocation and Controllable Sets}
\author{Suraj Kumar$^{1}$$^{*}$, Aditya Rallapalli$^{1}$, Ashok Kumar Kakula$^{1}$, Bharat Kumar GVP$^{1}$
\thanks{$^*$Corresponding Author}
\thanks{$^{1}$The authors are associated with Controls and Digital Area, U R Rao Satellite Center, Indian Space Research Organization, Bengaluru, Karnataka, India\{surajk, adityar, ashok, bharat\}@ursc.gov.in}
}

\begin{document}

\maketitle
\thispagestyle{empty}
\pagestyle{empty}

%%%%%%%%%%%%%%%%%%%%%%%%%%%%%%%%%%%%%%%%%%%%%%%%%%%%%%%%%%%%%%%%%%%%%%%%%%%%%%%%
\begin{abstract}
India achieved a significant milestone on August $23^{\text{rd}}$ 2023, becoming the fourth country to accomplish a soft landing on the Moon. This paper presents the powered descent trajectory design for the Chandrayaan-3 mission. The optimization framework is based on pseudospectral Radau collocation, and controllability-based waypoint refinement is employed to further enhance the robustness of the trajectory against state and control perturbations. Furthermore, the trade-off between fuel consumption and robustness is explicitly quantified, providing insights into the practical considerations of mission planning.
\end{abstract}
\keyword{\small Optimal Control, Trajectory Optimization, Lunar Landing, Pseudospectral Collocation, Controllable set}
%This paper presents a robust trajectory design framework for the powered descent phase of Chandrayaan-3, with specific emphasis on optimizing and refining guidance way-points to ensure safe and fuel-efficient lunar landing. The descent is divided into four key phases—Rough Braking, Attitude Hold, Fine Braking, and Terminal Descent—each governed by mission-specific constraints such as sensor operation windows, actuation limits, and navigation accuracy. A multi-phase optimal control problem is first solved offline to determine nominal way-points that minimize fuel consumption while satisfying phase-wise terminal constraints. To improve the robustness of these way-points against dispersions in state and actuation, a backward refinement process is proposed. This process employs sensitivity analysis and decision boundary classification in a reduced state space to identify and shift the fine braking start point deeper into the convergent region, while maintaining bounds on fuel usage and sensor operability. The refined trajectory is validated using a polynomial guidance law that solves a simplified two-point boundary value problem onboard. A two-stage optimization approach is formulated to maximize robustness with minimal increase in fuel consumption, ensuring that the final design remains feasible for onboard execution. The resulting trajectory exhibits improved tolerance to deviations, ensuring a more reliable soft landing under uncertainty.

%%%%%%%%%%%%%%%%%%%%%%%%%%%%%%%%%%%%%%%%%%%%%%%%%%%%%%%%%%%%%%%%%%%%%%%%%%%%%%%%
\section{INTRODUCTION}
The Moon’s proximity to Earth makes it a strategic gateway for deep-space missions. India’s Chandrayaan-1 discovered water on the lunar surface, sparking renewed global interest. Chandrayaan-2 attempted a soft landing in 2019 but was unsuccessful. Chandrayaan-3, launched as its successor, demonstrated autonomous landing and surface mobility, successfully touching down near the lunar south pole on August 23, 2023, at Shiv Shakti Point located at latitude $69.373560^\circ$S and longitude $32.319750^\circ$E.

Long-term and cost-effective lunar exploration is best achieved through autonomous missions employing a lander-rover combination. A major challenge in such missions is the requirement for the lander to perform a fully autonomous soft landing on the surface using only sensors, processors, and actuators on board, without human intervention. %To achieve mission objectives and ensure a precise landing sequence, the trajectory must satisfy multiple mission constraints. The challenge is further amplified when landing on an atmosphere-less body like the Moon, where the entire orbital velocity must be reduced using onboard propulsion alone. 
This demands not only a fuel-efficient trajectory but also one that is robust to mid-course deviations in nominal trajectory due to navigation and propulsion system perturbations. We emphasize on fuel efficiency, as the fuel optimal trajectory need not have the necessary robustness in adapting to state and control perturbations due to the usage of suboptimal analytical guidance laws in Chandrayaan-3. This implies that there is a trade-off between fuel optimality and robustness. Ensuring safety and mission success under such uncertainties makes robust trajectory design indispensable. 

Trajectory design involves optimizing a performance metric while satisfying state and control constraints. As a result, it is often formulated as an optimal control problem, also referred to as trajectory optimization. With the exception of simple problems such as infinite-horizon LQR, optimal control problems must be solved numerically. The need to solve optimal control problems numerically has led to a wide range of numerical approaches. These numerical approaches are broadly classified into two categories: indirect methods, which rely on solving the necessary conditions of optimality, and direct methods, which transcribe the optimal control problem into a nonlinear program. A survey of different methods for optimal control is provided in \cite{rao2009survey}. Collocation methods are among the most widely used direct approaches for trajectory optimization. Their popularity has grown with the rapid advancement of efficient solvers capable of handling large-scale nonlinear program\cite{betts1992application,byrd1999interior,vanderbei1999interior}. The two most common forms of collocation are local collocation and global collocation. Local collocation employs Runge-kutta methods\cite{dontchev2000second} and orthogonal collocation to discretize the optimal control problem\cite{reddien1979collocation}. The key difference between an orthogonal collocation and Runge-kutta method is the manner in which the collocation points are chosen. In orthogonal collocation, the collocation points are the roots of an orthogonal polynomial associated with a quadrature rule to approximate definite integrals. In recent years, pseudospectral methods have gained popularity. They are a global form of orthogonal collocation, where the state is approximated by a global polynomial, and collocation is applied at selected points. Unlike local collocation, which fixes the polynomial degree and varies the number of meshes, pseudospectral methods fix the number of meshes and vary the polynomial degree \cite{garg2010unified,garg2011direct}.

The main contribution of this work is the design of powered descent trajectory of Chandrayaan-3 mission. Unlike existing studies that focus solely on fuel-optimal trajectories, we show that such trajectories may lack robustness when realized by the onboard suboptimal guidance laws. To address this, we propose a two-stage design: the first stage uses global multiphase radau collocation to generate fuel optimal waypoints, while the second stage refines these waypoints using the controllable set of onboard guidance law to enhance robustness against state and control perturbations.

The paper is organized as follows. Section \ref{ocf} details the optimal control problem and Radau collocation formulation for trajectory optimization. Section \ref{pdtd} details the powered descent trajectory design and simulation results. Section \ref{sec:conc} concludes the paper.

\section{Optimal Control Formulation} \label{ocf}
\subsection{Multiphase Optimal Control Problem} 
The objective of an optimal control problem is to determine the state trajectory \(x(t) \in \mathbb{R}^n\), the control input \(u(t) \in \mathbb{R}^m\), the initial time \(t_0 \in \mathbb{R}\), and the terminal time \(t_f \in \mathbb{R}\), with \(t \in [t_0, t_f]\) as the independent variable, that minimizes the cost functional
\begin{equation}
\label{cost_fun}
J = \Phi\!\big[x(t_0), t_0, x(t_f), t_f\big] 
+ \int_{t_0}^{t_f} L\!\big[x(t), u(t), t\big] \, dt
\end{equation}
subject to the system dynamics
\begin{equation}
\dot{x}(t) = f\!\big[x(t), u(t), t\big],
\end{equation}
the path constraints
\begin{equation}
C_{\min} \leq C\!\big[x(t), u(t), t\big] \leq C_{\max},
\end{equation}
and the boundary conditions
\begin{equation}
\phi_{\min} \leq \phi\!\big[x(t_0), t_0, x(t_f), t_f\big] \leq \phi_{\max}
\end{equation}

In practice, many optimal control problems are divided into several phases, \(p \in \{1, \ldots, P\}\), which are then linked in a meaningful way. The overall cost is given as
\begin{equation}
J = \sum_{p=1}^{P} J^{(p)}
\end{equation}
where the phase cost \(J^{(p)}\) is given by Eq.~(\ref{cost_fun}).  
Each phase satisfies the phase specific dynamics
\begin{equation}
\dot{x}^{(p)}(t) = f\!\left(x^{(p)}(t), u^{(p)}(t), t\right),
\end{equation}
the path constraints
\begin{equation}
C_{\min}^{(p)} \leq C\!\left(x^{(p)}(t), u^{(p)}(t), t\right) \leq C_{\max}^{(p)},
\end{equation}
the boundary conditions
\begin{equation}
\phi_{\min}^{(p)} \leq \phi\!\left(x^{(p)}(t_0^{(p)}), t_0^{(p)}, x^{(p)}(t_f^{(p)}), t_f^{(p)}\right) 
\leq \phi_{\max}^{(p)},
\end{equation}
and the linkage constraints
\begin{equation}
\label{linkage_cnstr}
\begin{aligned}
L_{\min}^{(s)} \leq &\, L\!\big(x^{(l_s)}(t_f^{(l_s)}), u^{(l_s)}(t_f^{(l_s)}), t_f^{(l_s)}, \\
& \;\;\; x^{(r_s)}(t_0^{(r_s)}), u^{(r_s)}(t_0^{(r_s)}), t_0^{(r_s)}\big) 
\leq L_{\max}^{(s)}
\end{aligned}
\end{equation}
Here, \(r_s, l_s \in \{1, \ldots, S\}\) are the right and left phases, respectively, of the linkage pairs, with \(r_s \neq l_s\) (implying that a phase cannot be linked to itself) and \(S\) is the number of pairs of phases to be linked. While \cite{rao2009survey} considers more general linkage constraints, in this work we restrict them to sequential linkages, which implies which implies $r_s > l_s$.

\subsection{Pseudospectral Radau Collocation}
Pseudospectral collocation methods are best described in normalized time interval $\tau \in [-1,1]$. For notation simplicity, the superscript \(p\) denoting the phase is omitted up to Eq.~(\ref{lgr_dyn}).
For each phase, consider the following transformation:
\begin{equation}
t = \Big(\frac{t_f - t_0}{2} \Big)\, \tau + \Big(\frac{t_f + t_0}{2}\Big)
\end{equation}
Let us consider \( N \) Legendre-Gauss-Radau (LGR) collocation points \( \tau_1, \tau_2, \ldots, \tau_N \) on the interval \([ -1, 1 ]\), with \( \tau_1 = -1 \) and \( \tau_N < +1 \).
The LGR points are the roots of the the polynomial $P_{N-1}(\tau) + P_N(\tau)$ where $P_N(\tau)$ is the $N^{th}$ degree Legendre polynomial. LGR points have the property that
\begin{equation}
\begin{aligned}
    \int_{-1}^{+1} p(\tau) \, d\tau = \sum_{i=1}^{N} w_i \, p(\tau_i)
\end{aligned}
\end{equation}
is exact for polynomials of degree at most $2N-2$ where $w_i$ are the LGR quadrature weights. 

We introduce an additional non-collocated point \( \tau_{N+1} = 1 \) for discretization. The state trajectory of each phase is approximated by a polynomial of degree at most N as
\begin{equation}
    \boldsymbol{x}(\tau) \approx \boldsymbol{X}(\tau) = \sum_{i=1}^{N+1}\boldsymbol{X_i}L_i(\tau)
\end{equation}
where $\boldsymbol{X_i} \equiv \boldsymbol{X}(\tau_i)$ and $L_i$ is the Lagrange polynomial basis given by
\begin{equation}
    L_i(\tau) = \prod_{j=1,j\neq i}^{N+1} \dfrac{\tau-\tau_j}{\tau_i - \tau_j}, i=1,2,...,N+1
\end{equation}
Differentiating the state approximation and evaluating at the collocation point $\tau_k$ gives
\begin{equation}
\begin{aligned}
    \dot{\boldsymbol{X}}(\tau_k) = \sum_{i=1}^{N+1}\boldsymbol{X_i} \dfrac{dL_i(\tau_k)}{d\tau} = \sum_{i=1}^{N+1}D_{ki}\boldsymbol{X_i}    
\end{aligned}
\end{equation}
where $D_{ki} = \dot{L}_i(\tau_k)$. The derivative of the Lagrange polynomial at the collocated points is represented by the LGR differentiation matrix $D \in R^{N \times N+1}$. It has one row for each collocation point; the elements in the $i^{th}$ column are the derivatives of the Lagrange polynomial evaluated at each of the collocation points. 

The dynamics constraint for each phase is then transcribed into algebraic constraints as
\begin{equation}
\label{lgr_dyn}
    \begin{aligned}
    \sum_{i=1}^{N+1}D_{ki}\boldsymbol{X_i} - \frac{t_f-t_0}{2}f(X_k, U_k, \tau_k;t_0,t_f) = 0 \\
    k = 1, ..., N     
    \end{aligned}
\end{equation}
The original cost function is approximated using quadrature integral obtained from LGR points, 
\begin{equation}
    \begin{aligned}
        J =\; & \sum_{p=1}^{P} \Phi^{(p)}\left( X^{(p)}_0,\, t^{(p)}_0,\, X^{(p)}_f,\, t^{(p)}_f \right) +\\
             & \sum_{p=1}^{P} \frac{t^{(p)}_f - t^{(p)}_0}{2} 
        \sum_{k=1}^{N^{(p)}} w^{(p)}_k\, L^{(p)}\left( X^{(p)}_k,\, U^{(p)}_k,\, \tau^{(p)}_k \right)
    \end{aligned}
\end{equation}
where $N^{(p)}$ is the number of collocation points in phase p. Similarly the path constraints and boundary constraints are imposed at LGR points (including the non-collocated point at $\tau_{N+1} = 1$),
\begin{equation}
    \begin{aligned}
    C_{\min}^p \leq C(X_k^{(p)}, U_k^{(p)}, \tau_k^{(p)}; t_0, t_f) \leq C_{\max}^p \quad (k = 1, \ldots, N+1)
    \end{aligned}
\end{equation}
\begin{equation}
    \begin{aligned}
        \phi_{\min}^p \leq \phi_k(X_1^{(p)}, t_0, X_{N+1}^{(p)}, t_f) \leq \phi_{\max}^k
    \end{aligned}
\end{equation}
\begin{equation}
\begin{aligned}
L^{(s)}_{\min} \leq L^{(s)}\left( X^{(l_s)}_{N+1},\,U^{(l_s)}_{N+1}, t^{(l_s)}_f, X^{(r_s)}_0,\,U^{(r_s)}_0, t^{(r_s)}_0 \right) \leq L^{(s)}_{\max} \\
\quad l_s,\, r_s \in \{1, \ldots, S\}
\end{aligned}
\end{equation}
The state and control variable at each point is for phase is stacked in a column of decision vector, $\boldsymbol{z^{(p)}} \in \mathcal{R}^{n_x (N+1) + n_u N + 2}$ as
\begin{equation}
    \begin{aligned}
        \boldsymbol{z^{(p)}} \equiv \begin{bmatrix}
            \boldsymbol{z}_x, 
            \boldsymbol{z}_u,
            t_0,
            t_f
        \end{bmatrix}^{T}
    \end{aligned}
\end{equation}
where $\boldsymbol{z}_x$ is the vector of variables associated with the values of the state at the
discretization points, $\boldsymbol{z}_u$ is the vector of variables associated with the values of
the control at the LGR points, $t_0$ is the initial time, and $t_f$ is the terminal time associated with phase p. The decision variable ($\boldsymbol{z}$) for transcribed nonlinear program is then obtained by stacking decision variables of each phase. The resultant nonlinear program is given as 
\begin{equation}
\begin{aligned}
\min_{\boldsymbol{z}} \quad & f(\boldsymbol{z}) \\
\text{subject to} \quad & g(\boldsymbol{z}) = 0, \\
                        & h(\boldsymbol{z}) \leq 0
\end{aligned}
\end{equation}

The resultant nonlinear program is solved with MATLAB's fmincon wrapper using sequential quadratic program algorithm. 
\section{Powered Descent Trajectory Design} \label{pdtd}
%The powered descent trajectory of Chandrayaan-3 is initiated from initial parking orbit of 100x30 km (which is selected considering the orbit determination requirements). Broadly, orbital velocity is braked in two phases; first majority of horizontal velocity is reduced then followed by vertical velocity. The phases in the trajectory is further decided based on sensor operational constraints and safety during terminal descent phase.  

The powered descent trajectory of Chandrayaan-3 is broadly divided into four distinct phases, namely, Rough braking phase, Attitude hold phase, Fine braking phase and terminal descent phase. The Rough braking phase begins at 30 km altitude and extends till 7.4 km. During this phase, most of the initial velocity is reduced by firing four 800 N engines. This is followed by the attitude hold phase for absolute sensor operations. In this phase, the lander attitude is maintained 50$^{0}$ with respect to local vertical ensuring altimeters are pointed towards nadir. The attitude hold phase ends with sensor updates to inertial navigation system providing new initial states for next guidance phase. Fine braking, as the name suggests breaks the remaining velocities to near zero, brings the module to a safe altitude of 800 m above the designated landing site for terminal descent phase. The terminal descent phase involves vertical descent from 800 m altitude to touchdown. 

The task of trajectory optimization is to compute optimal waypoints for each phase while satisfying mission constraints. A two-stage design process is employed, considering sensor operation requirements, actuation limits, and inertial navigation errors to achieve fuel efficiency while maintaining robustness to state and control perturbations. The first stage performs a forward pass to generate a fuel optimal trajectory, followed by a backward pass that relaxes the waypoints to enhance robustness against state and propulsion dispersions. The resulting set of initial and terminal position and velocity waypoints for each phase are then realized by the onboard guidance law. Notably, the onboard guidance does not track the complete reference trajectory but utilizes only the initial and terminal states to generate reference accelerations by solving a two-point boundary value problem using polynomial guidance. % and/or fix additional trajectory parameters that are otherwise not included in trajectory optimization.

The attitude hold phase is an open-loop segment of 10s for sensor operation, during which the thrust vector is fixed at \(50^\circ\) with respect to vertical with constant magnitude. Since its state evolution can be obtained by direct propagation of the equations of motion, this phase is excluded from dynamic optimization. Consequently, trajectory optimization considers only the rough braking, fine braking, and terminal descent phases, with the initial condition of fine braking determined by propagating the equations of motion through the attitude-hold phase from the terminal state of rough braking.
\subsection{Forward Pass - Trajectory Optimization}
Trajectory optimization for the three phases is formulated as a constrained multiphase optimal control problem and solved using the Radau pseudospectral collocation method as discussed in Sec\ref{ocf}. The forward pass solution determines the optimal powered descent start point, total downrange to travel to the landing site, and trajectory waypoints for the onboard guidance laws. This reference trajectory also serves as a benchmark for quantifying the suboptimality gap obtained from onboard guidance law. We now describe the trajectory optimization formulation.
\iffalse
\begin{figure}[t]
  \centering
  \includegraphics[width=0.5\textwidth]{coordinate_frame.png}
  \caption{Coordinate Frame for Rough and Fine braking phase}
  \label{fig:coord_frame}
\end{figure}
\fi
The translational dynamics for rough braking and fine braking is considered in the moon-centered moon fixed frame using the spherical coordinate system as shown in Fig \ref{fig:coord_frame}. The 3D equations of motion is given by,
\begin{align}
\label{traj_opt_dyn}
    \dot{r} &= w \nonumber \\
    \dot{\theta} &= \frac{u}{rcos\phi} \nonumber \\
    \dot{\phi} &= \frac{v}{r} \nonumber \\
    \dot{w}  &= \frac{Tcos\alpha}{m} - \frac{\mu}{r^2}  + \frac{(u^2 + v^2)}{r} \nonumber \\ 
    &+ (2u\omega cos\phi + r\omega^2 cos^2\phi) \nonumber \\
    \dot{u} &= \frac{Tsin\alpha cos\beta}{m} + \frac{(-uw + uv tan\phi)}{r} \nonumber \\
    &+ (-2w\omega cos\phi + 2v\omega sin\phi) \nonumber \\
    \dot{v} &= \frac{(Tsin\alpha sin\beta)}{m} \frac{(-vw - u^2 tan\phi)}{r} \nonumber \\
    &+ (-2u\omega sin\phi - r\omega^2 sin\phi cos\phi) \nonumber \\
    \dot{m} &= -\frac{T}{I_{sp}g_0}
\end{align}
Here, $r$ represents the radial distance from the moon centre, $\theta, \phi$ represents the lander latitude and longitude respectively; $\omega$ represents moon rotational velocity; $v,u,w$ represents the tangential, across and radial velocity components respectively; $m$ represents the mass of lander; $\mu$ represents moon's gravitational constant; $I_{sp}$ represents specific impulse of the engine and $g_0 = 9.81$. The thrust vector is parameterized in the body frame by its magnitude $T$, in-plane pitch angle $\alpha$ and out-of-plane yaw angle $\beta$. 

In terminal descent phase, lander dynamics is switched to local terrain relative frame defined by north, east and up direction as shown in Fig \ref{fig:local_frame}. The dynamics for the terminal descent phase is therefore appropriately expressed in this frame as at low altitudes and reduced velocities, the Moon can be approximated as a non-rotating flat body. Under this assumption, terminal descent phase dynamics is given as,
\begin{eqnarray}
&\dot{\boldsymbol{{r}}} &=\quad \boldsymbol{{v}} \nonumber \\
&\boldsymbol{\dot{{v}}} &=\quad \boldsymbol{{g}} + \boldsymbol{{a}} \nonumber \\
&||{a}|| &=\quad \frac{T}{m}   \nonumber \\
&\dot{m} &=\quad -\frac{T}{I_{sp} g_{0}}  \label{lander_dynamics4}
\end{eqnarray}

where $\boldsymbol{r} \in \mathcal{R}^3$ , $\boldsymbol{v} \in \mathcal{R}^3$, $\boldsymbol{a} \in \mathcal{R}^3$ and $\boldsymbol{g} \in \mathcal{R}^3$ are defined with respect to local frame. 

The global trajectory is segmented into multiphase trajectory composed of rough braking, fine braking, and terminal descent phase. %As noted earlier, the terminal descent phase is further divided into multiple sub-phases, including multiple hovering maneuvers, re-targeting, and vertical descent, all aimed at ensuring the lander's safety. The waypoints for these sub-phases are determined solely based on the lander’s altitude and are not part of the trajectory optimization process. Instead, the reference trajectory generated through trajectory optimization serves as a benchmark to evaluate the trade-offs in optimality introduced by incorporating multiple hovering maneuvers—where the lander's velocity is first brought to zero, then re-accelerated, and again reduced to zero. Although such operations consume additional fuel, they provide significant safety margins. The trajectory optimization assumes a single-shot landing during the terminal descent phase and yields the minimum fuel requirement, which can be used to quantify the additional fuel needed to enhance safety. 
Let $x^{(i)}$ denote a variable $x$ in phase i = \{1, 2, 3\}. The objective of the rough braking phase is to reduce maximal orbital velocity and ensure that lander horizontal velocity is favorable for Lander Position Detection Camera (which provides lander absolute position with respect to moon surface) at sufficient high altitude in order to provide absolute sensor information to navigation which otherwise operate only with inertial sensors in rough braking phase. The targetted altitude cannot be targetted too high as camera imaging resolution is poor but can also not be planned as too low otherwise it will be too late to correct errors accumulated in inertial navigation. Therefore, altitude of 7.4 km is chosen as terminal altitude for rough braking phase. 

The initial condition for fine braking is given by the terminal condition of rough braking, propagated for 10s with a thrust vector at an angle of 50$^\circ$ with respect to the vertical and constant magnitude. Fine braking reduces remaining velocity at the end of terminal velocity to zero, and bring the module vertically up above the landing site. %Table  \ref{fb_boundary_condition} lists the boundary constraints of Fine braking phase. 
Finally, terminal descent phase brings module from the altitude of 800 m to surface with zero velocity. %Table \ref{vd_boundary_condition} lists the boundary constraints of Fine braking phase.
\iffalse
\begin{equation}
\begin{aligned}
\text{Initial:} \quad 
&r_0^{(3)} = r_f^{(2)}, \quad 
\theta_0^{(3)} = \theta_f^{(2)}, \quad 
\phi_0^{(3)} = \phi_f^{(2)}, \\
&w_0^{(3)} = w_f^{(2)}, \quad 
u_0^{(3)} = u_f^{(2)}, \quad 
v_0^{(3)} = v_f^{(2)}, \\
&m_0^{(3)} = m_f^{(2)}, \quad 
\alpha_0 = 0, \quad 
\beta_0 = 0
\end{aligned}
\end{equation}

\begin{equation}
\begin{aligned}
\text{Final:} \quad 
&r_f^{(3)} = R_M, \quad 
\theta_f^{(3)} = \theta_{PLS}, \quad 
\phi_f^{(3)} = \phi_{PLS}, \\
&w_f^{(3)} = 0, \quad 
u_f^{(3)} = 0, \quad 
v_f^{(3)} = 0, \\
&m_f^{(3)} = \text{Free}, \quad 
\alpha_f = 0, \quad 
\beta_f = 0
\end{aligned}
\end{equation}

\begin{table}[ht]
\centering
\begin{tabular}{|c|c|c|c|}
\hline
\textbf{Phase} & \textbf{States} & \textbf{Initial} & \textbf{Final} \\
\hline
3 & r & $r_{0}^{(3)} = r_{f}^{(2)}$ & $r_{f}^{(3)} = R_{M}$ \\
        & $\theta$ & $\theta_0^{(3)} = \theta_f^{(2)}$ & $\theta_f^{(3)} = \theta_{PLS}$ \\
        & $\phi$  & $\phi_0^{(3)} = \phi_f^{(2)}$  & $\phi_f^{(3)} = \phi_{PLS}$ \\
        & w & $w_{0}^{(3)} = w_{f}^{(2)}$  & $w_{f}^{(3)} = 0$ \\
        & u & $u_{0}^{(3)} = u_{f}^{(2)}$  & $u_{f}^{(3)} = 0$ \\
        & v & $v_{0}^{(3)} = v_{f}^{(2)}$  & $v_{f}^{(3)} = 0$ \\
        & m & $m_{0}^{(3)} = m_{f}^{(2)}$ & $m_{f}^{(3)} = $ Free \\
        & $\alpha$ & 0 & 0 \\
        & $\beta$  & 0 &  0 \\
\hline
\end{tabular}
\caption{Terminal Descent Boundary Constraints}
\label{vd_boundary_condition}
\end{table}
\fi

Eq (\ref{bc_summary}) shows the boundary constraints for each phase. Free variables are not shown. 
\begin{equation}
\label{bc_summary}
\begin{aligned}
&\textbf{Phase 1 (Rough Braking):} \\
&r_0^{(1)} = R_M + 30~\text{km}, \quad r_f^{(1)} = R_M + 7.4~\text{km}, \\
&w_0^{(1)} = 0, \; u_0^{(1)} = 0, \; v_0^{(1)} = 1.68~\text{km/s}, \; v_f^{(1)} < v_{sops}, \\
&m_0^{(1)} = 1729~\text{kg}, \; \alpha_f^{(1)} = 50^\circ, \; \beta_f^{(1)} = 0 \\
&\textbf{Phase 2 (Fine Braking):} \\
&r_0^{(2)} = \mathcal{P}^{AH}(r_f^{(1)}), \quad r_f^{(2)} = R_M + 800~\text{m}, \\
&\theta_{0}^{(2)} = \mathcal{P}^{AH}(\theta_f^{(1)}), \quad \theta_{f}^{(2)} = \theta_{PLS}, \\
&\phi_{0}^{(2)} = \mathcal{P}^{AH}(\phi_f^{(1)}), \quad \phi_{f}^{(2)} = \phi_{PLS}, \\
&w_0^{(2)} = \mathcal{P}^{AH}(w_f^{(1)}), \quad u_0^{(2)} = \mathcal{P}^{AH}(u_f^{(1)}), \quad v_0^{(2)} = \mathcal{P}^{AH}(v_f^{(1)}), \\
&w_f^{(2)} = 0, \; u_f^{(2)} = 0, \; v_f^{(2)} = 0, \\
&\alpha_0^{(2)} = 50^\circ, \; \alpha_f^{(2)} = 0, \; \beta_0^{(2)} = \beta_f^{(2)} = 0, \\
&m_0^{(2)} = \mathcal{P}^{AH}(m_f^{(1)}) \\
&\textbf{Phase 3 (Terminal Descent):} \\
&\textbf{r}_0^{(3)} = [0,r_f^{(2)}-R_M,0]^T, \quad \textbf{r}_f^{(3)} = [0,0,0]^T \\
&\textbf{v}_0^{(3)} = [v_f^{(2)},w_f^{(2)},u_f^{(2)}]^T, \quad \textbf{v}_f^{(3)} = [0,0,0]^T \\
\end{aligned}
\end{equation}
Here $R_M$ represents the radius of the moon; $\theta_{PLS},\phi_{PLS}$ represents the latitude and longitude of the landing site respectively; $v_{sops}$ denotes the horizontal velocity threshold for sensor operation.

\begin{figure}[htbp]
\centering
%---------------------- (a) ----------------------
\begin{subfigure}{0.48\columnwidth}
\centering
\tdplotsetmaincoords{70}{120}
\begin{tikzpicture}[scale=1.2,tdplot_main_coords]
% Moon as circle
\draw[thick] (0,0,0) circle (1);
% Axes
\draw[->] (0,0,0) -- (1.5,0,0) node[below]{$X$};
\draw[->] (0,0,0) -- (0,1.5,0) node[right]{$Y$};
\draw[->] (0,0,0) -- (0,0,1.5) node[above]{$Z$};
% Position vector r
\tdplotsetcoord{P}{10}{60}{40}
\draw[thick,->] (0,0,0) -- (P) node[midway, above=3pt]{$r$};
\fill[red] (P) circle (1.5pt);

\coordinate (Rp) at (Pxy); % Pxy = (P_x, P_y, 0)
\coordinate (Pxy) at ({1.2*cos(40)*cos(60)},{1.2*cos(40)*sin(60)},0);
\draw[dashed] (0,0,0) -- (Pxy);
% Angles
\tdplotdrawarc[->]{(0,0,0)}{0.5}{0}{60}{anchor=north}{$\theta$} % 0 → XY-projection
\foreach \t in {0,5,...,40} {
    \pgfmathsetmacro{\X}{0.9*cos(\t)*cos(60)}
    \pgfmathsetmacro{\Y}{0.9*cos(\t)*sin(60)}
    \pgfmathsetmacro{\Z}{0.9*sin(\t)}
    \coordinate (A\t) at (\X,\Y,\Z);
}
\draw[->,black] (A0) -- (A5) -- (A10) -- (A15) -- (A20) -- (A25) -- (A30) -- (A35) -- (A40)
node[right] {$\phi$};
% Local spacecraft axes
\draw[->,dashed] (P) -- ++(-1,-1.,0) node[below]{$x$};
\draw[->,dashed] (P) -- ++(-0.4,0.4,0) node[right]{$y$};
\draw[->,dashed] (P) -- ++(1,1,1) node[above]{$z$};
\end{tikzpicture}
\caption{Spacecraft orbiting the moon}
\end{subfigure}
%---------------------- (b) ----------------------
\begin{subfigure}{0.48\columnwidth}
\centering
\tdplotsetmaincoords{70}{120}
\begin{tikzpicture}[scale=1.2,tdplot_main_coords]
% Body axes
\draw[->] (0,0,0) -- (1.5,0,0) node[below]{$x,u$};
\draw[->] (0,0,0) -- (0,1.5,0) node[right]{$y,v$};
\draw[->] (0,0,0) -- (0,0,1.5) node[above]{$z,w$};
\tdplotsetcoord{T}{4}{55}{40}
\draw[thick,->] (0,0,0) -- (T) node[above right] {$T$};
% XY-plane projection
\coordinate (Tp) at ({2*cos(40)*cos(50)},{2*cos(40)*sin(50)},0);
\draw[dashed] (0,0,0) -- (Tp); % optional
% Alpha: XY-plane angle from x-axis to projection of T
\tdplotdrawarc[->]{(0,0,0)}{0.4}{0}{50}{anchor=north}{$\beta$}
\tdplotsetthetaplanecoords{40} % align plane with T's azimuth
\tdplotdrawarc[tdplot_rotated_coords,->,thick]
  {(0,0,0)}{0.6}{0}{55}{pos=0.1,above right}{$\alpha$}

\end{tikzpicture}
\caption{Thrust vector in body frame}
\end{subfigure}
\caption{Coordinate Frame for Rough and Fine braking phase}
\label{fig:coord_frame}
\end{figure}
The objective of trajectory optimization is the minimization of propellant consumption, which, equivalently, is posed as the maximization of vehicle mass at the final time in the final phase, 
\begin{equation}
    J = -m(t_f^{(3)})
\end{equation}
The control constraints are given as 
\begin{eqnarray}
   &T_{min} \leq T \leq T_{max} \label{thrust_constraint1}
\end{eqnarray}

\begin{figure}
\begin{subfigure}{0.1\columnwidth}
\centering
\begin{tikzpicture}[scale=0.8, every node/.style={font=\sffamily}]
  % ---------- Styles ----------
  \tikzset{
    axis/.style={very thick, -{Latex[length=3.5mm]}},
    traj/.style={blue!70!black, dashed, line width=0.8pt},
    attitudeAxis/.style={very thick, -{Latex[length=3.5mm]}, color=blue!70!black}
  }

  % ---------- Moon surface and local frame ----------
  \draw[line width=0.8pt, gray] (-0.5,-0.0) -- (3.5,-0.0) node[black, right, xshift=6mm]{Moon surface};
  \coordinate (O) at (2,0);

  % Local frame

    \draw[->] (O) -- ++(-2.2,0) node[below]{+North}; 
    \draw[->] (O) -- ++(1.6,1.0) node[right]{+East}; 
    \draw[->] (O) -- ++(0,3.0) node[right]{+Up};

  % ---------- Trajectory ----------
  \draw[traj] plot [smooth,tension=1] coordinates { (0.4,4.5) (1.1,3.8) (1.7,3.0) (2.1,2.0) (2.0,0.6) };

  % ---------- Lander (rectangle) ----------
  \coordinate (L) at (1.0,3.9);
  \draw[fill=gray!20, draw=black, thick,rotate=25] ($(L)+(-0.2,-0.1)$) rectangle ++(0.3,0.6);

  % ---------- Attitude axes ----------
  \draw[->,rotate=25] (L) -- ++(-1.0,0) node[left]{+$x$};
  \draw[->,rotate=25] (L) -- ++(0.0,1.0) node[above right]{+$y$};
  \node[rotate=-20, anchor=west] at ($(L)+(0.3,0.2)$) {+$z$ (Inside the plane)};

  % ---------- Origin dot ----------
  \fill (O) circle (0.7pt);
\end{tikzpicture}
\end{subfigure}
\caption{Coordinate frame for terminal descent phase}
  \label{fig:local_frame}
\end{figure}

Eq.(\ref{thrust_constraint1}) is associated with the physical limits on the thrust by the engines respectively. Upto fine braking phase, 4 engines are operational and in terminal descent phase, two engines are switched off. To account for engine underperformance which cannot be measured in flight, trajectory optimization is carried out considering the maximum allowable specification in the underperformance ($5\%$ of maximum rating of 800 N) in each engine.

\begin{figure}[hbt!]
  \centering
  \begin{minipage}{.23\textwidth}
    \centering
    \includegraphics[width=0.9\textwidth]{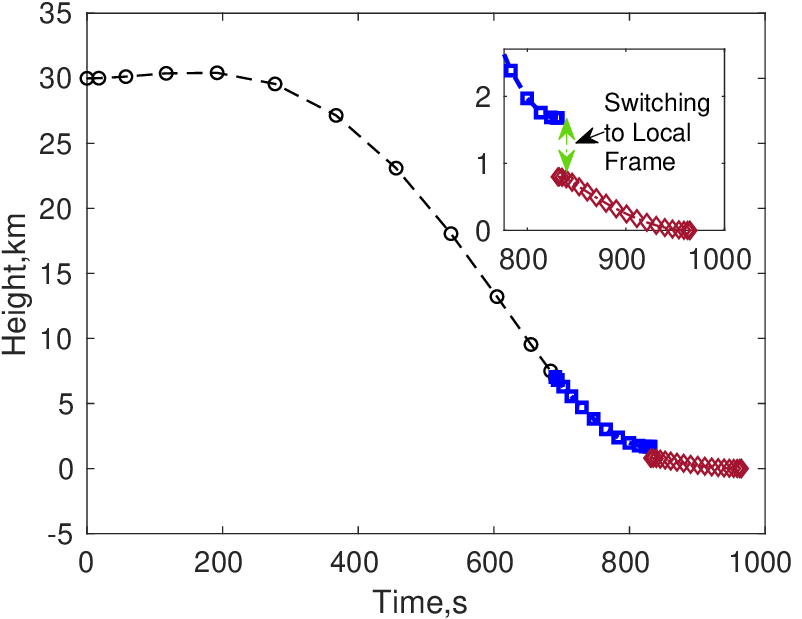}
    \subcaption{Altitude}\label{fig:fopdg_alt}
  \end{minipage}
  \begin{minipage}{.23\textwidth}
    \centering
    \includegraphics[width=0.9\textwidth]{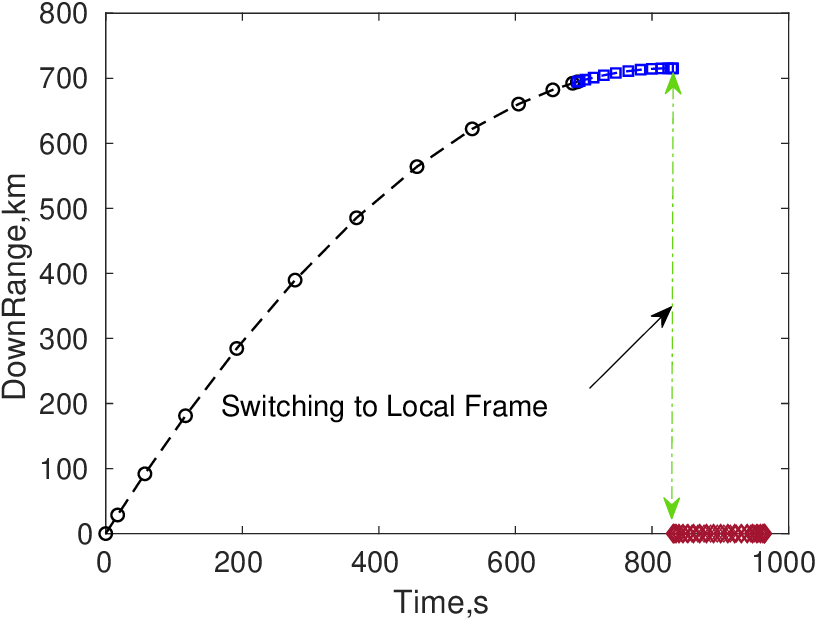}
    \subcaption{Downrange}\label{fig:fopdg_dr}
  \end{minipage}
  \begin{minipage}{.23\textwidth}
    \centering
    \includegraphics[width=0.9\textwidth]{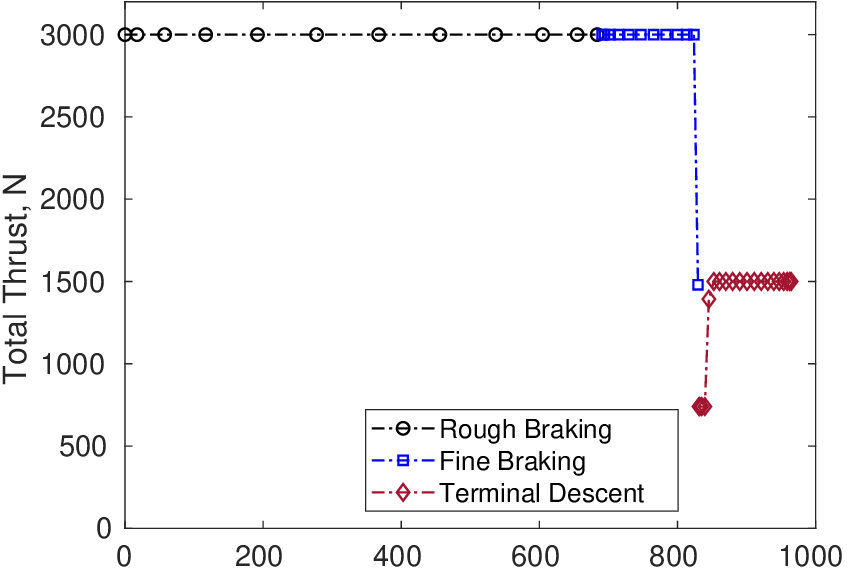}
    \subcaption{Thrust Per Engine}\label{fig:fopdg_thr}
  \end{minipage}
    \begin{minipage}{.23\textwidth}
    \centering
    \includegraphics[width=0.9\textwidth]{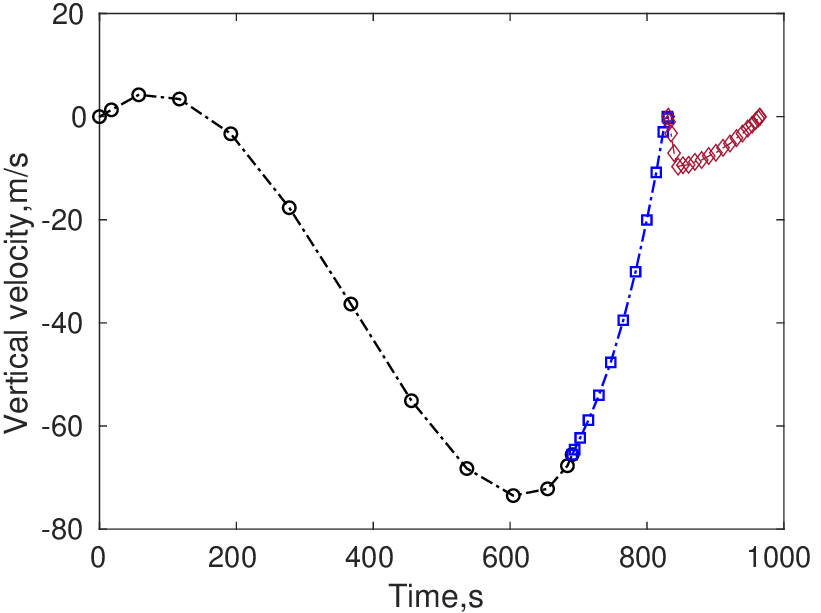}
    \subcaption{Vertical Velocity}\label{fig:fopdg_vv}
  \end{minipage}
  \begin{minipage}{.23\textwidth}
    \centering
    \includegraphics[width=0.9\textwidth]{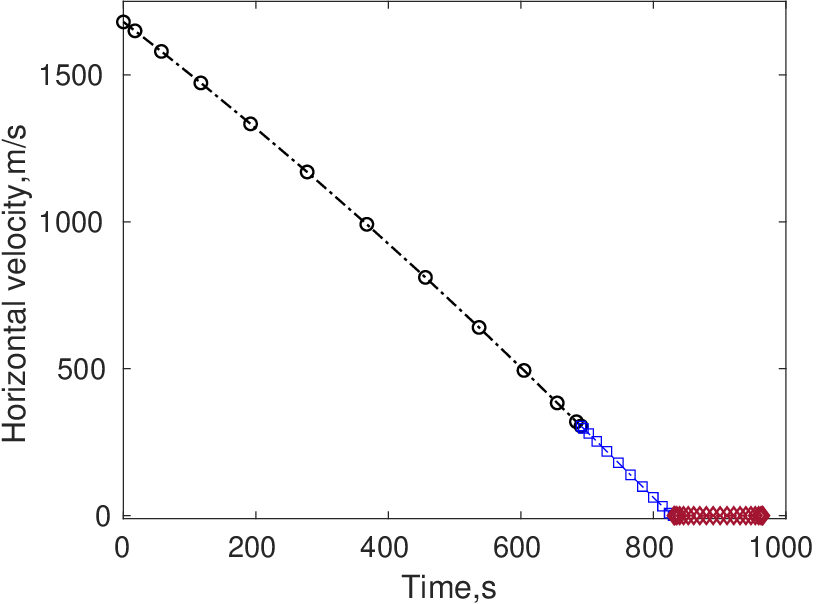}
    \subcaption{Horizontal Velocity}\label{fig:fopdg_hv}
  \end{minipage}
  \begin{minipage}{.23\textwidth}
    \centering
    \includegraphics[width=0.9\textwidth]{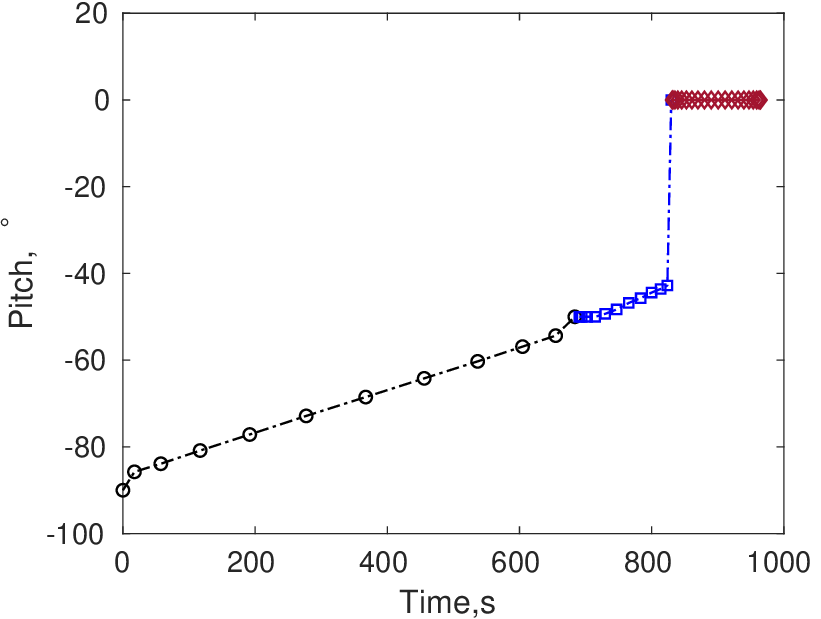}
    \subcaption{Pitch Angle}\label{fig:fopdg_awv}
  \end{minipage}
  \caption{Fuel Optimal Powered Descent Trajectory}
  \label{fig:fopdg_traj}
\end{figure}

Figure~\ref{fig:fopdg_traj} illustrates the state and control trajectories obtained from the multi-phase trajectory optimization. Figures~\ref{fig:fopdg_alt} and~\ref{fig:fopdg_dr} show the optimized altitude and downrange profiles for the entire powered descent. The discontinuities observed in altitude and downrange occur due to changes in reference frame and associated dynamics. Up to the fine braking phase, altitude is measured from the mean lunar surface, whereas in the local frame used thereafter, it is measured relative to the local terrain, which is elevated by 883 meters above the mean lunar surface. Similarly, the downrange resets to zero as the origin of the local frame is fixed at the ground projection of the realized terminal state of the fine braking phase. The thrust profile shown in Figure~\ref{fig:fopdg_thr} exhibits a bang-bang structure, characteristic of minimum fuel formulation.  An interesting observation in the pitch profile of the lander ( in-plane angle) shows that it is fuel optimal to perform a quick pitch up maneuver at the end of fine braking phase to reorient the lander for vertical descent. Since the de-orbit maneuver is designed such that the landing site lies directly beneath the orbital track, no out-of-plane corrections are required. As a result, the crossrange and lateral velocity remain effectively zero. %It is also observed that all phase-specific constraints are satisfied by the optimization. Because the horizontal velocity constraint at the end of the rough braking phase is specified as an inequality, the optimizer selects the terminal velocity that minimizes fuel consumption while respecting the constraint. %Table \ref{tab:rb_fb_states} shows the optimized way-points obtained from trajectory optimization. 
%\begin{table}[h!]
%\centering
%\caption{Optimized Way-points}
%\label{tab:rb_fb_states}
%\begin{tabular}{|c|c|c|c|}
%\hline
%\textbf{Phase} & \textbf{Quantity} & \textbf{Initial Value} & \textbf{Terminal Value} \\
%\hline
%\multirow{6}{*}{RB} 
%& Height (km)            & $30$   & $7.4$   \\
%& Downrange (km)         & $-715$  & $-24$  \\
%& Crossrange (m)        & $0$  & $0$  \\
%& Vertical Velocity (m/s)   & $0$  & $-65$  \\
%& Horizontal Velocity (m/s) & $1680$ & $305$ \\
%& Cross Velocity (m/s)      & $0$ & $0$ \\
%\hline
%\multirow{6}{*}{FB} 
%& Height (km)            & $7.4$   & $1.68$   \\
%& Downrange (km)         & $-24$  & $0$  \\
%& Crossrange (km)        & $0$  & $0$  \\
%& Vertical Velocity (m/s)   & $-65$  & $0$  \\
%& Horizontal Velocity (m/s) & $305$ & $0$ \\
%& Cross Velocity (m/s)      & $0$ & $0$ \\
%\hline
%\end{tabular}
%\end{table}

Finally, the waypoints obtained from trajectory optimization are implemented through a suitable guidance algorithm in real time. Perhaps, the most widely cited approach for solving the fuel-optimal powered descent guidance problem is lossless convexification \cite{malyuta2022convex}, which transforms the original non-convex fuel optimal control problem into a convex optimization problem that can be solved onboard in real time. Although such methods explicitly handle constraints, the computational burden on low-speed processors limits their real-time applicability. Consequently, onboard guidance laws typically rely on analytical formulation under simplifying assumptions, often neglecting state and actuation constraints, and thus yield suboptimal guidance law. Therefore, it is essential to evaluate the robustness of the suboptimal guidance law when applied to the optimized waypoints. This necessitates a backward pass in the trajectory design process, wherein the waypoints are further refined by taking into account the specific characteristics and limitations of the onboard guidance law.
\subsection{Backward Pass - Controllability based Refinement}
In this stage, the waypoints obtained from trajectory optimization are further refined to enhance robustness against state and control dispersions, while balancing the trade-off with fuel consumption. We present a generic strategy of waypoint refinement using the characterization of controllable set in reduced state space. Although applicable across all phases, we demonstrate its use in the fine braking phase, where robustness is critical since its initial conditions can vary significantly due to navigation updates from absolute sensors in the preceding attitude hold phase.

We first describe the flight guidance algorithm and characterize its controllable set, which forms the key element for the backward pass. The Chandrayaan-3 guidance algorithm is based on classical polynomial guidance which was significantly improved through simulation based optimization to improve robustness and fuel efficiency. To retain analytical formulation, the translational dynamics is considered under the flat-moon assumption similar to terminal descent phase, given by Eq.~(\ref{lander_dynamics4}). Let $\boldsymbol{x} = [\mathbf{r}^T,\mathbf{v}^T]^T$ denotes the state vector. The waypoints for guidance are specified in terms of initial and terminal position, velocity and acceleration, respectively.
\begin{equation}
\label{boundary_conditions}
\begin{aligned}
    \mathbf{r}(t_0) = \mathbf{r}_0, \quad 
    \mathbf{v}(t_0) = \mathbf{v}_0, \quad 
    \mathbf{a}(t_0) = \mathbf{a}_0, \\
    \mathbf{r}(t_f) = \mathbf{r}_f, \quad 
    \mathbf{v}(t_f) = \mathbf{v}_f, \quad 
    \mathbf{a}(t_f) = \mathbf{a}_f.
\end{aligned}
\end{equation}
Acceleration waypoints are imposed to ensure continuity across phase while position and velocity waypoints are obtained from the the trajectory design process explained in this paper. The net acceleration ($\mathbf{\bar{a}} = \mathbf{a} + \mathbf{g}$) is represented by a third-order polynomial in time:
\begin{equation}
\mathbf{\bar{a}}(t) = \mathbf{C}_{0} + \mathbf{C}_{1}t + \mathbf{C}_{2}t^2 + \mathbf{C}_{3}t^3
\end{equation}
The coefficients of the polynomial are arrived by solving the systems of linear equations,
\begin{equation}
\label{gui_lin_eq}
    \begin{bmatrix} 
    1 & 0 & 0 & 0 \\
    1 & t_{\text{go}} & t_{\text{go}}^2 & t_{\text{go}}^3 \\
    t_{\text{go}} & \frac{t_{\text{go}}^2}{2} & \frac{t_{\text{go}}^3}{3} & \frac{t_{\text{go}}^4}{4} \\
    \frac{t_{\text{go}}^2}{2} & \frac{t_{\text{go}}^3}{6} & \frac{t_{\text{go}}^4}{12} & \frac{t_{\text{go}}^5}{20} \\
\end{bmatrix}\begin{bmatrix} C_{0}^i \\ C_{1}^i \\ C_{2}^i \\ C_{3}^i \end{bmatrix} = \begin{bmatrix}
\boldsymbol{a}_0^i \\
\boldsymbol{a}_f^i \\
\boldsymbol{v}_f^i - \boldsymbol{v_0}^i \\
\boldsymbol{r}_f^i - \boldsymbol{r_0}^i - \boldsymbol{v_0}^i \cdot t_{\text{go}} \\
\end{bmatrix} \\
\end{equation}
where $i \in {x,y,z}$ and $t_{go} = t_f -t$ denotes the remaining time to reach the target. In each guidance cycle, the net acceleration is computed by substituting the current states of the lander as the initial states to determine the polynomial coefficients. The free parameter $t_{go}$ is obtained through simulation based optimization in a supervised learning framework. It is modeled as a policy mapping of the state—downrange ($S$), altitude ($H$), vertical velocity ($w$), and horizontal velocity ($v$)—to the fuel-optimal $t_{go}$ within the class of functions parameterized by cubic polynomial.

%Although not commonly emphasized in the literature, this secondary refinement step significantly improves the trajectory's robustness—particularly in missions like Chandrayaan-3 that use onboard sub-optimal analytical guidance laws, which do not explicitly solve an optimal control problem under state and control constraints. 
%Specifically, the trajectory design is refined in a backward manner, starting from the landing site and proceeding up to the initiation of powered descent. First, the nominal start way-point of the Fine Braking phase is adjusted using linear parametric perturbations around the optimized trajectory to enhance the dispersion-handling capability of the onboard polynomial guidance law. Once the Fine Braking start point is finalized, the terminal point of the Rough Braking phase is determined by back-propagating the system state through the Attitude Hold phase. Finally, the initial point of the Rough Braking phase is fine-tuned to complete the refinement process.

\begin{figure}[h]
  \centering
  \includegraphics[width=0.4\textwidth]{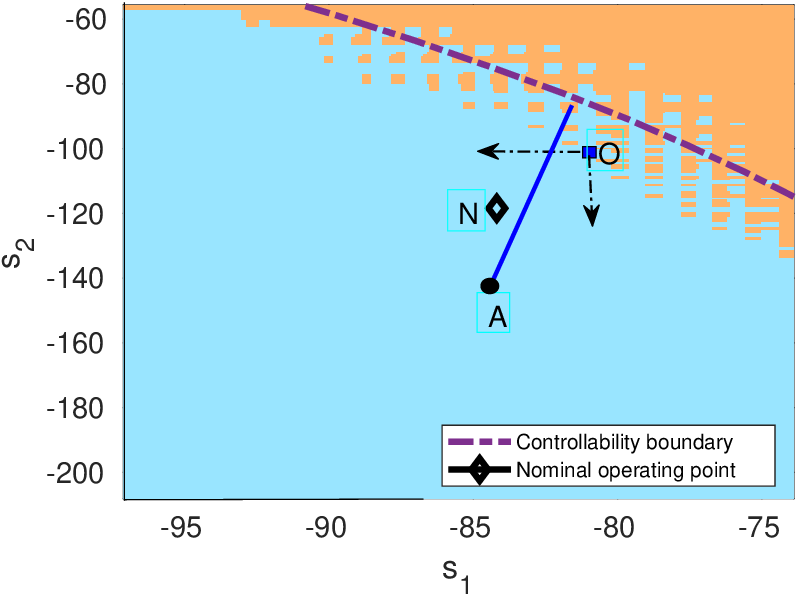}
  \caption{Controllable set and Refinement of fine braking initial waypoint. Point O denotes the waypoint obtained from trajectory optimization. Point N denotes the relaxation in the waypoint to maximize robustness}
  \label{fig:guidance_data_set}
\end{figure}

\begin{figure}[h]
  \centering
  \includegraphics[width=0.4\textwidth]{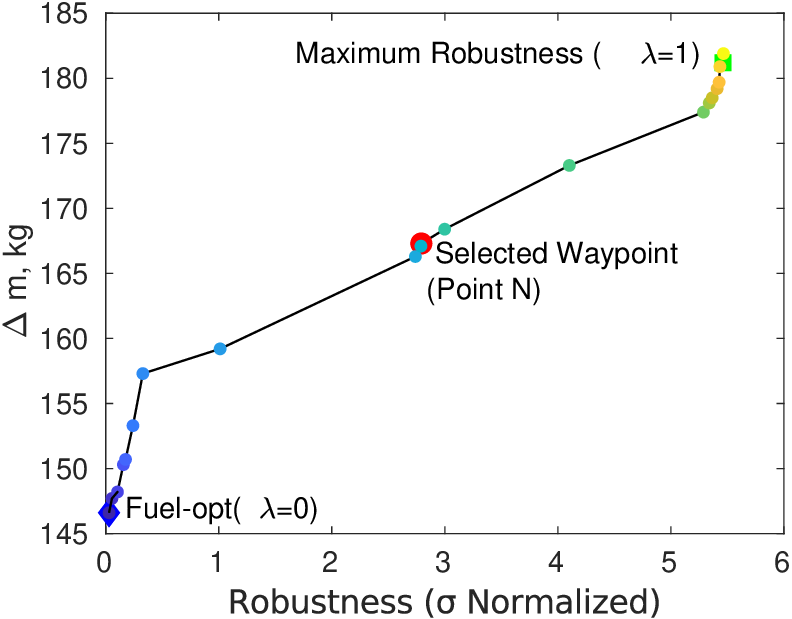}
  \caption{Fuel Optimality vs Robustness Tradeoff}
  \label{fig:tradeoff}
\end{figure}

%\subsubsection*{Fine Braking Refinements}
\iffalse
The polynomial guidance law can be concisely represented as a closed-loop operator,
\begin{equation}
\mathbf{a} = \mathcal{G}(\mathbf{r}_0, \mathbf{v}_0, \mathbf{a}_0, \mathbf{r}_f, \mathbf{v}_f, \mathbf{a}_f, t_{\text{go}})
\end{equation}
\fi
Let $\boldsymbol{\bar{x}}_0$ be the optimized initial state vector obtained from the forward pass and $\Sigma_0$ the covariance matrix representing the allowable dispersions in the initial state vector. We construct the guidance dataset $\Xi$ as
\begin{equation}
\begin{aligned}
    \Xi := \{\boldsymbol{\bar{x}}_0 + \delta_i \;|\; \delta_i \sim \mathcal{N}(0, \Sigma_0), \; i = 1,2,\ldots,|\Xi| \}
\end{aligned}
\end{equation}
For each $\boldsymbol{x}_i \in \Xi$, the cubic polynomial guidance is rolled out over $t_{go} \in [t^{\min}_{go}, t^{\max}_{go}]$, yielding the set
\begin{equation}
    \mathcal{F}(\boldsymbol{x}_i) = \{ (t_{go,k}, m_{f,k}) \;|\; k = 1,2,\ldots,|\mathcal{F}(\boldsymbol{x}_i)| \}
\end{equation}
where $m_f$ denotes the final mass in that trajectory rollout. Each element of this set represents the tuple of $t_{go}$ and terminal mass for the given state vector. The fuel-optimal $t_{go}^*$ and propellant usage for each state are
\begin{equation}
\begin{aligned}
    t_{go}^*(\boldsymbol{x}_i) = \arg\max_{(t_{go}, m_f) \in \mathcal{F}(\boldsymbol{x}_i)} m_f, \\
    \Delta m^*(\boldsymbol{x}_i) = m_0 - m_f(t_{go}^*; \boldsymbol{x}_i)
\end{aligned}
\end{equation}
where $m_0$ is the mass at the start of the phase. The supervised learning dataset is then given by
\begin{equation}
    \mathcal{D} = \{ (\boldsymbol{x}_i, t_{go}^*(\boldsymbol{x}_i), \Delta m^*(\boldsymbol{x}_i)) \;|\; i = 1,2,\ldots,|\Xi| \}
\end{equation}
This dataset is used to construct a surrogate model for both the fuel-optimal $t_{go}$ and corresponding propellant consumption. Finally, the controllable set is given by the union of all nonempty sets $\mathcal{F}(\boldsymbol{x}_{i})$:
\begin{equation}
\label{cont_set}
    \mathcal{X}_C := \bigcup_{x_{i} \in \Xi} \mathcal{F}(x_{i})
\end{equation}
Using the controllable set $\mathcal{X}_c$, the classification dataset can be created as 
\begin{equation}
\label{eq:cont_dataset}
\mathcal{T} = \{ (\boldsymbol{x}_i, d_i) \mid \boldsymbol{x}_i \in {\Xi}, d_i \in \{-1, +1\} \}
\end{equation}
where $d_i$ is the label, with $d_i = +1$ denoting controllable states (for which polynomial guidance converges to the desired terminal conditions) and $d_i = -1$ denoting uncontrollable states. This classification dataset can now be used for waypoint refinement. Let the reduced state vector be defined as
\[
\mathbf{s} = \begin{bmatrix} s_1 \\ s_2 \end{bmatrix} =
\begin{bmatrix} S/v \\ H/w \end{bmatrix}
\] 
Fig \ref{fig:guidance_data_set} shows the dataset in this reduced state space. The point labeled O represents the optimized fine braking waypoint obtained from the forward pass. Since dispersion in optimized state is assumed to follow a Gaussian process, the controllability boundary is parameterized using a convex conic function using maximum margin classifier \cite{chakrabarti2024convex} as
\begin{equation}
    \partial\mathcal{X}_C(s_1, s_2) = a s_1^2 + b s_1 s_2 + c s_2^2 + d s_1 + e s_2 + f = 0
\end{equation}
where $a, b, c, d, e, f$ are the coefficients obtained from maximum-margin classifier on the classification dataset. As evident, the optimized waypoint lies near the controllability boundary. To enhance robustness, it can be shifted within the controllable region while keeping the altitude fixed to meet sensor constraints and adjusting the downrange and velocity within allowable limits. However, this introduces a tradeoff—moving deeper into the controllable region increases dispersion tolerance but also fuel consumption. A practical robustness metric is defined as the signed distance of the selected waypoint from the decision boundary:
\begin{equation}
    \mathcal{M}(s_1^A, s_2^A) = 
    \frac{\partial\mathcal{X}_C(s_1^A, s_2^A)}{\|\nabla \partial\mathcal{X}_C(s_1^A, s_2^A)\|}
\end{equation}
A positive value of $\mathcal{M}(s_1^A, s_2^A)$ indicates that waypoint $A$ lies within the controllable region, ensuring convergence to the terminal state, while a negative value implies non-convergence. The magnitude $|\mathcal{M}(s_1^A, s_2^A)|$ approximates the orthogonal distance of $A$ to the decision boundary and quantifies the robustness. The robust fine braking waypoint is obtained by simultaneously maximizing robustness and minimizing fuel consumption using a weighted-sum formulation:
\begin{equation}
\min_{S,v,w} \; \mathcal{J}(S,v,w) = -\lambda \mathcal{M}(S/v,H/w) + (1-\lambda) \Delta m^*(S,H,v,w)
\end{equation}
subject to operational constraints:
\[
v \le v_{\mathrm{sops}}, \quad \mathcal{M}(S/v,H/w) > 0, \quad \Delta m(S,H,v,w) \le \Delta m_{\max}
\]
Even though the controllable region is convex, $\Delta m^*$ is not monotonic, therefore, multiple points may achieve the same maximal value of $\mathcal{J}$. To select a single waypoint, the solution with minimum propellant usage among all maximal points is chosen. Fig.~\ref{fig:tradeoff} shows the trade-off curve between robustness and fuel optimality, where robustness is normalized by its standard deviation. By varying $\lambda \in [0, 1]$, the trade-off curve is traced. The curve can be interpreted as the increase in fuel consumption per unit improvement in robustness (measured in $\sigma$ units). For the mission, robustness is maintained at the $3\sigma$ level, with the corresponding waypoint $N$ shown in Fig.~\ref{fig:guidance_data_set}. Clearly, an additional 25 kg of propellant is utilized to achieve this level of robustness.

Once the fine braking start point (point $N$) is finalized, the waypoint for the rough braking phase is obtained by back-propagating the fine braking start state through the attitude hold phase. The terminal altitude of the rough braking phase remains unchanged, as it is not included in the parametric variation discussed earlier. Consequently, the only remaining variable to be determined is the downward travel during the rough braking phase. The start point of the rough braking phase is determined using a one-dimensional line search over downrange, wherein the downrange is varied from its minimum to maximum value. A similar trade-off curve is utilized to maintain robustness near the midpoint of the trade-off curve. However, detailed results are omitted here due to space limitations.
\subsection{Terminal Descent Trajectory Design}
We briefly describe the terminal descent phase strategy here. While the trajectory optimization assumes a single vertical descent from an altitude of 800 m to the surface, the actual terminal descent phase is further divided into multiple subphases to enable terrain-relative navigation, hazard detection, and avoidance for a safe and soft landing. These subphases include multiple hovering segments for camera-based operations. 

\begin{figure}[h]
\centering
\includegraphics[width=0.48\textwidth]{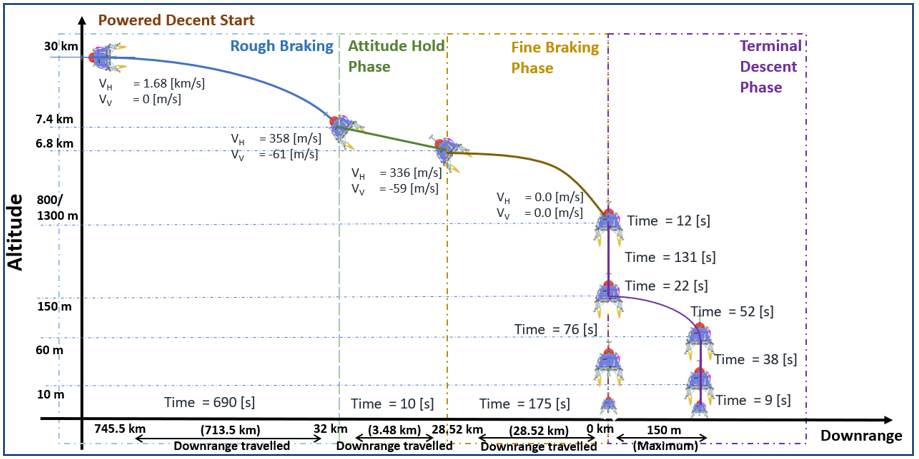}
\caption{Powered Descent Trajectory of Chandrayaan-3 }\label{pdphase}
\end{figure}

The terminal descent phase starts with 12 sec hovering where two engines are shut down to ensure guidance required thrust is within engines operational range. lander navigation is switched from inertial reference to terrain relative frame centered below the lander on the local terrain. At the end of the first hovering, guidance uses the latest navigation estimates to plan the vertical descent to 150 m. Vertical descent phase ends at 150 m altitude with a planned second hovering phase for hazard detection. The second hovering phase lasts for a maximum of 22 seconds. During this phase, the hazard camera is operated to provide hazard-free coordinates with respect to current location. Once a hazard free landing point is identified, hovering is terminated and retargeting to safe location is initiated with latest states estimated from navigation. Retargeting is done to an altitude of 60 m above hazard free landing location ensuring collision free retargeting. After retargeting, the lander descends to 10 meters, aiming for a target vertical velocity of -1 m/s, followed by a constant velocity of -1 m/s descent until touchdown sensor detects surface contact with footpad. Fig. \ref{pdphase} shows the various waypoints obtained after the trajectory design for the final flight. More details on the landing profile and powered descent trajectory for Chandrayaan-3 can be found in \cite{rallapalli2024landing}.

%\subsubsection*{Rough Braking Refinements}

%the rough braking start point is relaxed to improve the dispersion handling capacity of onboard sub-optimal analytical guidance law. This is done by relaxing the downrange traveled to the landing site. This further ensured that the onboard rough braking guidance remains robust against perturbations of up to 48 N in each engine (even when way-points are designed to cater for under-performance of 37N each engine). Similarly, the fine braking start waypoint is modified to improve the kinematic conditions in the fine braking phase while ensuring the sensor constraints in Eq.(\ref{sensor_cnstr}) are met. Finally, we fix the desired initial and terminal acceleration in each phase that is not part of trajectory optimization to maintain acceleration continuity across the phases. Although not frequently reported in the literature, this second pass significantly improves the trajectory's robustness under state vector and propulsion system dispersions and improves the safety margins. Its effect is more pronounced when using analytical sub-optimal onboard guidance law which doesn't explicitly solve an optimal control problem under state and action constraints such as that of Chandrayaan-3 sub-optimal onboard guidance law. Fig. \ref{pdphase} shows the various waypoints obtained after the trajectory design for the final flight. More details on the landing profile and powered descent trajectory for Chandrayaan-3 can be found here. \cite{rallapalli2024landing} 

\section{Conclusion}  \label{sec:conc}
This work presented a systematic approach to powered descent trajectory design for Chandrayaan-3, incorporating both optimality and robustness considerations. A two-stage design process was incorporated, wherein fuel-optimal way-points are first generated through trajectory optimization and subsequently refined using controllable set and robustness metrics tailored to the onboard sub-optimal guidance law. The resulting trajectory ensured safe and fuel-efficient landing performance while enhancing tolerance to dispersions in state and control.
\section{Acknowledgment}
The authors would like to express sincere gratitude to Chairman ISRO for his continuous support. Authors would also like to thank Mr. Sankaran M (Director URSC), Mr. Sudhakar S (Deputy Director, Controls and Digital Area), Dr. Veeramuthuvel (Project Director, Chandrayaan - 3) and Dr. U P Rajeev (Group Director MSSG, VSSC) for their guidance and suggestions during Chandrayaan-3 key reviews.

\bibliography{sample.bib}

\end{document}